\documentclass[seceq]{ptptex}

\usepackage{graphicx}
\usepackage{wrapft}

\title{QCD Sum Rules and 1/$N_c$ expansion}

\author{Toru \textsc{Kojo}$^1,$ and Daisuke \textsc{Jido}$^2$,
}

\inst{$^1$RBRC, Brookhaven National Laboratory, Upton,
	NY 11973-5000, USA\\
$^2$YITP, Kyoto University, Kyoto 606-8502, Japan}

\abst{The 1/$N_c$ arguments are developed to classify 
the hadronic states in the correlators.
Arguments applied to the $\sigma$ meson correlator enable
to separate the instanton, glueball, and, in particular,
the $\pi\pi$ scattering states by $1/N_c$ from
both 2q and 4q correlators.
The bare resonance pole with no mixing effects are analyzed
with the QCD sum rules (QSR).
The results suggest the existence of nontrivial correlation for
the mass reduction of 4q system.
}

\begin{document}

\maketitle

\section{$1/N_c$ classication of hadronic states}
The QCD based description of the 
exotic hadrons is one of the current problems
to study the nonperturbative properties of QCD
beyond those manifested in the usual hadron spectra.
The $\sigma$ meson with
$I=J=0$ is a typical example with
almost all important ingredients of the exotics.
The $\sigma$ meson could include not only usual 
$q\bar{q}$ (2q) state but hadronic states beyond the simple
constituent quark picture:
glueball, $\pi\pi$ molecule, and $qq\bar{q}\bar{q}$ (4q) states,
and is good laboratory to study the properties of 
these hadronic states and their interplay \cite{sigma}.
The studies of the $\sigma$ meson, however,
are not straightforward. Since it can be described 
as admixture of several hadronic states,
it is difficult to identify which hadronic states
are responsible to which part of the $\sigma$ meson properties.
Therefore, it is important to theoretically clarify the properties of
each hadronic state without mixing effects.

For this purpose, we introduce the $1/N_{c}$ classifications
of hadronic states in the correlators,
${\rm \Pi}(q) = \int d^{4}x \, e^{iq \cdot x}
\langle T J(x) J^{\dagger}(0) \rangle$.
One of the largest virtues of $1/N_{c}$ expansion 
\cite{tHooft,Witten}
is that it relates the quark-gluon dynamics to
the hadronic states order by order in terms of $1/N_{c}$,
reflecting their qualitative difference through the
difference in $1/N_c$ counting.
We will identify and separate 
the instanton, glueball, and, in particular,
$\pi\pi$ scattering states,
to concentrate on the remaining bare "2q" and "4q" states
(concrete definition will be given later)
without the complication from the mixing effects.
The "2q" and "4q" are dynamically studied through the
QCD sum rules (QSR) \cite{Shifman} which has
the firm theoretical ground in large $N_c$ limit.

The operators used in the correlators are as follows: 
The 2q interpolating fields are described as
$J_{M}^{F}=\bar{q} \tau_F \Gamma_M q$,
where Dirac matrix $\Gamma_M =(1, \gamma_{\mu})$
and $\tau_F\ (F=1,2,3)$ are the Pauli matrices 
acting on $q=(u,d)^T$.
The 4q operators with the $\sigma$ quantum number
are given (assuming the ideal mixing for the $\sigma$ meson)
by $J_{MM}(x) = \sum_{F=1}^3 J_{M}^{F}(x) J_{M}^{F}(x)$
as products of meson operators.

Here we give the inductive definition of "2q" and "4q"
states based on the $1/N_c$ counting for 
the 2q and 4q correlators.
Since
the large $N_c$ 2q correlators are known to have
the 2 meson scattering, glueball, and instanton only in the
quark-gluon graphs of higher order of $1/N_c$, 
we can regard the leading $N_c$ contributions 
as the sum of the bare "2q" states \cite{tHooft, Witten}.
Next we consider the 4q correlator,
incorpolating the 4q participating diagrams
from the beginning.
Here we give the inductive definition of the
"4q" state following the $1/N_c$ based orthogonal
condition:
a) "4q" is {\it not}
generated in the leading $N_{c}$ 2q correlator,
b) "4q" can appear in the 4q correlators
even in the absence of the 2-meson states.
Thus the dynamical origin of "4q" is different
from "2q" and 2-meson molecule states.
The glueball and instanton are easily verified
as the higher order effects in $1/N_{c}$ than those
we will consider,
thus will not be discussed further in the following.

\begin{figure}[t]
\vspace{-0.2cm}
\begin{center}
\hspace{0.7cm}
\includegraphics[width=13.0cm, height=2.8cm]{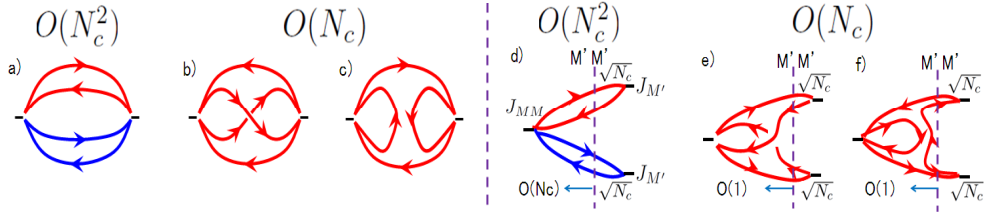} 
\vspace{-0.2cm}
\caption{Examples of the
$O(N_c^2)$ and $O(N_c)$ quark-gluon diagrams  
for 2 and 3 point correlators.}
\vspace{-1.0cm}
\label{fig:2pointgraph}
\end{center}
\end{figure}

The studies of the "4q" component in the $\sigma$ meson
require systematic steps
beyond the leading $1/N_{c}$ arguments for the 4q correlators,
since leading order $O(N_{c}^2)$ 
quark-gluon diagrams include only 2 planar loops 
(Fig.\ref{fig:2pointgraph},a),
which are interpretted as irrelevant free 2-meson scattering.
Thus we must proceed to the next leading order of $1/N_{c}$,
$O(N_c)$ diagrams
which could include the "2q" and "4q" states. 
To discuss 2-point correlators in compact description,
we classify the overlap strength
of the operator $J_{MM}$ with hadronic states by $1/N_c$.
The explicit examination of quark-gluon diagrams in
the 3-point correlator among $J_{MM}$ 
and two separated $J_{M'}$
(Fig.\ref{fig:2pointgraph}, d-f) indicates that
the overlap strength of 4q field 
with 2 meson states is
$\langle 0|J_{MM}|M'M'\rangle = O(N_c)\delta_{MM'} + O(1)+....$
since the leading order diagrams are $O(N_c^2)$ ($O(N_c)$) for $M=M'$
($M \neq M'$) and
the overlap strength of $J_M'$ with 2q meson state $|M'\rangle$ 
is $O(N_{c}^{1/2}$).
On the other hand,
the overlap strength with "2q" and "4q" states
cannot be deduced from $1/N_c$ arguments only,
and are assumed to be
$\langle 0|J_{MM}|R\rangle = O(N_c^{1/2}), 
\ \ (R="2q"\ {\rm or}\ "4q")$
which will be assured after the dynamical calculations.
Following our $1/N_c$ orthogonal condition a),
"4q" will be identified
by examining the quantitative difference 
of poles in 2q and 4q correlators,
which is found to be large enough 
to distinguish the "4q" and "2q" states.
The coupling of $R$ 
to two mesons is estimated by 3-point function in the same way 
to obtain the meson couplings performed in Ref.~\cite{Witten} and 
is found to be $O(N_{c}^{-1/2})$.

Now we can classify
the hadronic states in
2-point correlators $\langle J_{MM} J_{M'M'} \rangle$ 
based on $1/N_{c}$ (See, Fig.\ref{fig:2pointgraph}, a-c):
(i) If $M=M'$, $O(N_c^2)$ quark-gluon graphs include only the free
2M scattering states in the region $E \ge 2m_M$.
Otherwise, the contributions from these quark-gluon diagrams vanish,
indicating the absence of 2 meson scattering states.
(ii) $O(N_c)$ graphs include the 2M or 2M' scattering
and possible resonance, "2q" and/or "4q".
Note that in the case of
$M,M'\neq P\ {\rm or}\ A$,
2$\pi$ scattering states are not included up to $O(N_c)$ diagrams,
and then the resonance peaks (if exist) {\it below} $2m_{M}$
are isolated and have no width since
the decay channels are absent.
Therefore,
now we can reduce the $\sigma$ spectrum in the 4q correlator
into peak(s) plus continuum  
{\it if we retain only diagrams up to $O(N_c)$}.

\section{QCD Sum Rules for the reduced spectra}

This separate investigation of the $O(N_c^2)$ 
and $O(N_c)$ part enables to perform
the step by step analyses for
the each hadronic state.
In the application of the QSR,
we perform the operator product expansion (OPE)
for the correlators in deep Euclidean region ($q^2=-Q^2$),
then relate them, {\it term by term of $1/N_c$},
to the {\it integral} of the hadronic 
spectral function through the
dispersion relation:
\begin{eqnarray}
{\rm \Pi}^{ope}_{N_c^n}(-Q^2) 
= \int_0^{\infty}\hspace{-0.2cm}
ds\ \frac{1}{\pi}\frac{ {\rm Im \Pi}^h_{N_c^n}(s) }{s+Q^2}
\ \ (n=2,1).
\label{disp}
\end{eqnarray}

Now we emphasize the practical aspects of
$1/N_c$ expansion in the application of the QSR.
First, the higher dimension condensates in the OPE,
whose values have not been well-known
despite of their importance, can be factorized into
the products of known condensates,
$\langle \bar q q \rangle$, $\langle G^2 \rangle$, and
$\langle \bar q g_s \sigma G q \rangle$.
To keep this merit,
we will deduce the final $O(N_c)$ results from
the off diagonal correlator $\langle J_{VV}J^{\dag}_{SS} \rangle$,
whose leading order is $O(N_c)$
thus without the factorization violations
at the $O(N_c)$ OPE. 

Secondly, the lowest resonance in the reduced
$O(N_c)$ spectra ${\rm Im \Pi}_{N_c}^{h}(s)$
for "2q" and "4q" states,
can be described as the sharp peak 
because of the absence of the decay channel.
Applying the usual quark-hadron duality approximations
to the higher excited states,
$\pi{\rm Im \Pi}^{h}_{N_c} (s) = \lambda^2 \delta(s-m_h^2)
+ \theta(s-s_{th}) \pi {\rm Im \Pi}_{N_c}^{ope} (s)$,
and after the Borel transformation for the Eq.(\ref{disp}),
we can express the effective mass as
\begin{eqnarray} 
m_h^2(M^2;s_{th}) \equiv \frac{ \int_0^{s_{th}} \!ds
\ e^{-s/M^2 }s\ {\rm Im \Pi}^{ope} (s) }
 { \int_0^{s_{th}} \!ds\ e^{-s/M^2 }{\rm Im \Pi}^{ope} (s) }.
 \label{eq:peakmass}
\end{eqnarray}
$s_{th}$ can be uniquely fixed to satisfy the
least sensitivity \cite{Shifman} of the expression 
(\ref{eq:peakmass})
against the variation of $M$,
since the physical peak should not depend on the
unphysical expansion parameter $M$.
This criterion is justified only when the peak
is very narrow,
and our $1/N_c$ reduction of spectra 
is essential for its application
to allow the QSR framework to determine all physical parameters
($m_h, \lambda, s_{th}$)
in self-contained way.

In practical application of the QSR,
it is essential 
to reduce the error of the finite order trunction of the OPE
and of the quark-hadron duality approximation.
Thus Eq.(\ref{eq:peakmass}) must be treated in the window
($M^2_{min},\ M^2_{max}(s_{th})$)
to achieve the conditions: good OPE convergence for $M_{min}$
(highest dimension terms $\le$ 10\% of whole OPE)
and sufficient ground state saturation for $M_{max}$ 
(pole contribution $\ge$ 50\% of the total) \cite{KHJ,KJ}.
Without $M^2$ constraint,
we are often stuck with the {\it pseudo-peak} artifacts
\cite{KJ} often seen in the multiquark SRs.
Thus we carry out the OPE up to dimension 12
to find the reasonable $M^2$ window
\cite{KJ,KHJ}.

The $N_c$ dependence of the quantities
entering the OPE is summarized as follows.
The gauge coupling constant behaves like $O(N_c^{-1/2})$, 
and the condensates, $\langle O \rangle=$($\langle \bar{q}q\rangle$,
$\langle \alpha_s G^2 \rangle$, 
$\langle \bar q g_s \sigma G q \rangle$)
are $O(N_c)$.
Here we additionaly put the simple $N_c$ 
scaling assumption on these values,
$\alpha_s |_{N_c} = 3\alpha_s/N_c$,
$\langle O \rangle |_{N_c} = \langle O \rangle N_c/3$.
We take the values with errors for $N_c=3$ case,
$\alpha_s({\rm 1GeV}) =0.4$,
 $\langle \alpha_s G^2/\pi \rangle = (0.33\ {\rm GeV})^4$,
$\langle \bar{q}q\rangle=-(0.25 \pm 0.03\ {\rm GeV})^3$,
and 
$m_0^2 = \langle \bar q g_s \sigma G q \rangle/ 
\langle \bar{q}q\rangle
= (0.8 \pm 0.1)\ {\rm GeV^2}$, 
respectively.
The results shown below will use
the central value.

\begin{figure}[t]
\vspace{-0.3cm}
  \begin{center}
    \begin{tabular}{ccc}
      \resizebox{60mm}{!}{\includegraphics{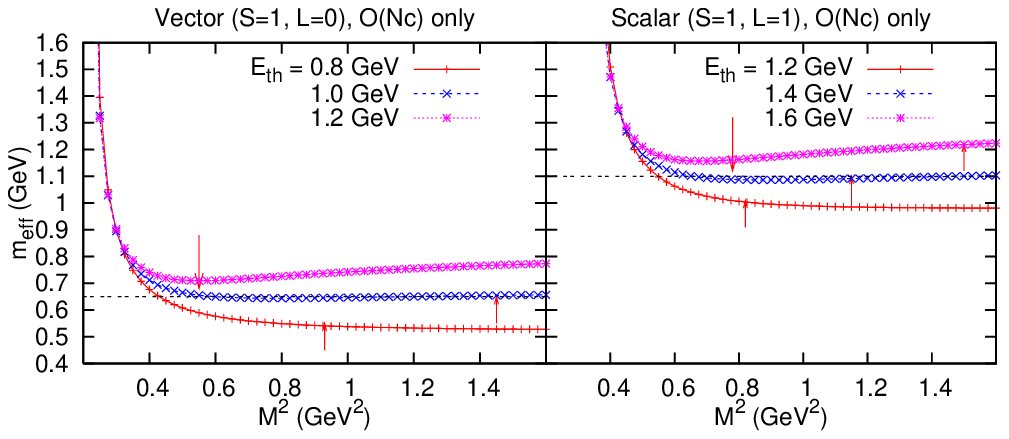}} &
      \resizebox{35mm}{!}{\includegraphics{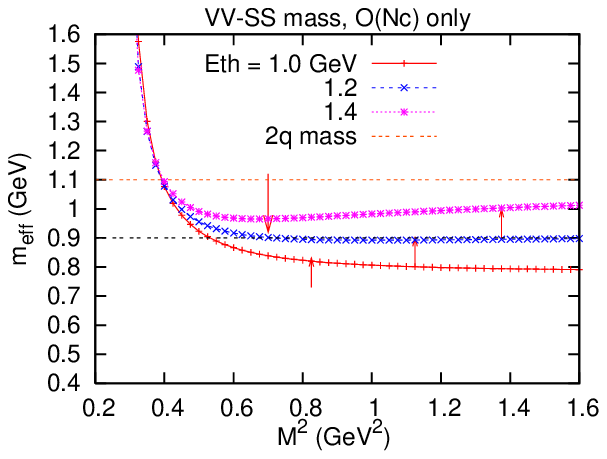}} &
	\hspace{-0.5cm}
      \resizebox{35mm}{!}{\includegraphics{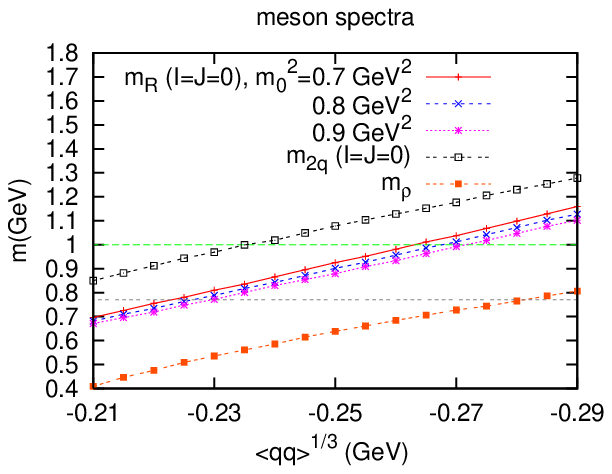}} \\
    \end{tabular}
	\vspace{-0.2cm}
    \caption{The large $N_c$ 2q correlator results for
   the scalar and vector mesons (left),
	the $O(N_c)$ part of the 4q correlators (middle),
 and their mass relation for the various condensate values
(right).}
	\label{fig:borelfig}
  \vspace{-0.4cm}
  \end{center}
\end{figure}

First we show in the left panel of Fig.\ref{fig:borelfig}
the results of
the large $N_c$ 2q correlators (expanded up to dimension 6)
for vector meson as a reference and scalar meson as the "2q" state
in the $\sigma$ meson.
The downarrow (upperarrow) indicates the values of 
$M^2_{min}$ ($M^2_{max}(s_{th})$).
Following the $E_{th}$ ($\equiv \sqrt{ s_{th} }$) fixing criterion,
we fix $E_{th}$ to 1.0 (1.4) GeV for vector (scalar) mesons,
and determine the mass as 0.65 (1.10) GeV.

Now we turn to the $O(N_c)$ part of
the 4q correlator,
$\langle J_{VV}J^{\dag}_{SS} \rangle$,
to investigate "2q" and "4q" states.
Shown in the middle panel of Fig.\ref{fig:borelfig}
are the effective masses
for $E_{th}$=1.0, 1.2, and 1.4 GeV.
We select the $E_{th}$=1.2 GeV case
and evaluate its mass as $\sim$0.90 GeV,
which is obviously lower than that of "2q" scalar meson case, 
$\sim$1.10 GeV 
in large $N_c$ limit,
and thus is considered as the mass of "4q" state.
We have also investigated
$\langle J_{SS}J^{\dag}_{SS} \rangle$
($\langle J_{VV}J^{\dag}_{VV} \rangle$),
and obtained the almost same mass 0.80 (0.90) GeV
although they could suffer from the factorization violation
coming from $O(N_c^2)$ OPE.
This 3-independent correlator analyses consistently suggest 
the existence of "4q" state lighter than "2q" state.
Finally, in the right panel of Fig.\ref{fig:borelfig},
we show the $\langle \bar{q}q \rangle$
and $m_0^2$ dependence of
"2q" vector, scalar meson masses,
and of the "4q" mass deduced from 
$\langle J_{VV}J^{\dag}_{SS} \rangle$. 
The inequality $m_{\rho}<m_{4q}<m_{S}$ holds
irrespective of the condensate values.

In conclusion,
the $1/N_c$ expansion
is useful to classify the hadronic states in the correlators,
especially of the multiquark operators.
The qualitative difference between several
hadronic states is related to the difference in $1/N_c$ counting.
A novel consequence of this approach is that
we can investigate the {\it bare} 4q state which is generated from the
genuine 4q dynamics, not factorized into 
the 2-meson dynamics.
The results obtained here
suggest that the $\sigma$ meson has the 
"4q" component, which includes the nontrivial correlation
needed for the considerable mass reduction \cite{KJ2}.

We thank for the organizers of {\it New Fronteers in QCD}
held at Yukawa Institute of Theoretical Physics (YITP).
This work is supported in part by the 
Grant for Scientific Research (No. 20028004)
in Japan.

\end{document}